\documentclass[aip,
 amsmath,amssymb,
 reprint,
]{revtex4-1}

\usepackage{graphicx}
\usepackage{dcolumn}
\usepackage{bm}
\usepackage{siunitx} 
\usepackage[utf8]{inputenc}
\usepackage[T1]{fontenc}
\usepackage{mathptmx}
\usepackage{etoolbox}

\makeatletter
\def\@email#1#2{%
 \endgroup
 \patchcmd{\titleblock@produce}
  {\frontmatter@RRAPformat}
  {\frontmatter@RRAPformat{\produce@RRAP{*#1\href{mailto:#2}{#2}}}\frontmatter@RRAPformat}
  {}{}
}
\makeatother
\begin{document}

\preprint{AIP/123-QED}

\title[State Selective Preparation and Nondestructive Detection of Trapped O$_2^+$]{State Selective Preparation and Nondestructive Detection of Trapped O$_2^+$}
\author{Ambesh Pratik Singh}
\author{Michael Mitchell}%
\author{Will Henshon}
\author{Addison Hartman}
\author{Annika Lunstad}
\author{Boran Kuzhan}
\author{David Hanneke$^*$}
 \email{dhanneke@amherst.edu}
\affiliation{ 
$^1$Department of Physics \& Astronomy, Amherst College, Amherst, Massachusetts 01002, USA
}%

\begin{abstract}

The ability to prepare molecular ions in selected quantum states enables studies in areas such as chemistry, metrology, spectroscopy, quantum information, and precision measurements.
Here, we demonstrate $(2+1)$ resonance-enhanced multiphoton ionization (REMPI) of oxygen, both in a molecular beam and in an ion trap. The two-photon transition in the REMPI spectrum is rotationally resolved, allowing ionization from a selected rovibrational state of O$_2$. Fits to this spectrum determine spectroscopic parameters of the O$_2$ $d\,^1\Pi_g$ state and resolve a discrepancy in the literature regarding its band origin. 
The trapped molecular ions are cooled by co-trapped atomic ions. Fluorescence mass spectrometry nondestructively demonstrates the presence of the photoionized O$_2^+$.
We discuss strategies for maximizing the fraction of ions produced in the ground rovibrational state. For $(2+1)$ REMPI through the $d\,^1\Pi_g$ state, we show that the Q(1) transition is preferred for neutral O$_2$ at rotational temperatures below 50~K, while the O(3) transition is more suitable at higher temperatures. 
The combination of state-selective loading and nondestructive detection of trapped molecular ions has applications in optical clocks, tests of fundamental physics, and control of chemical reactions.

\end{abstract}

\maketitle

\section{Introduction\protect\\}

Quantum state control of trapped molecular ions has broad applications. 
State-prepared molecular ions allow studies of quantum effects in chemical reactions, such as reactive resonances, $s$-wave reactive scattering, and quantum tunneling in low barrier potential wells.\cite{dashevskaya2003low,krems2005molecules,willitsch2008chemical} While the thermal motion of molecules precludes coherent control of collisions at room temperature,\cite{krems2008cold} state-prepared molecular ions have been used to implement controlled chemical reactions\cite{PhysRevA.73.042712} and to study reaction dynamics.\cite{bell2009ion,dorflerNatureComm2019} More recently,  translationally and internally cooled ions have been used to study increasingly complex chemical reactions that are likely to occur in extraterrestrial environments such as the interstellar medium, planetary atmospheres, and comets.\cite{schmid2020isomer,catani2020translationally}

In metrology, high-precision measurements of vibrational transitions determine the proton-to-electron mass ratio at the $10^{-11}$ level of fractional uncertainty.~\cite{patraScience2020,kortunovNaturePhys2021} Other vibrational transitions are candidates for clocks at near-IR and optical wavelengths.\cite{schillerPRL2014,carolloAtoms2018,wolfNJP2024} Such vibrational clocks could make sensitive tests of physical laws and  probe for variation of fundamental constants.\cite{hannekeQST2021} Trapped molecular ions provide the leading constraint on the existence of the electron's electric dipole moment,\cite{roussyScience2023} with molecules looking to play a key role in future searches.\cite{angPRA2022,arrowsmithKronRPP2024}

The variety of applications comes in part because of molecules' rotational and vibrational degrees of freedom. However, preparing these states of molecular ions can be complicated. Many of them are slow to decay or are prone to excitation by blackbody radiation or collisions. A cryogenic reduction of the ambient temperature is a longstanding technique for reducing the number of populated states. Inelastic collisions with a buffer gas can reduce the internal temperature of trapped ions. This has been used with a variety of molecular ions~\cite{gerlichPhysScripta1995} and proposed for use with nonpolar ions, whose rotation and vibration do not couple strongly to blackbody radiation.\cite{schillerPRA2017} Co-trapping molecular ions with laser-cooled atomic ions leads to translational cooling, but the interaction is sufficiently long-range that it does not impact the molecules' internal state.~\cite{bertelsenJPB2006}

Recent techniques inspired by those used in quantum information can couple the internal states of trapped molecules and atoms. Quantum projection can then prepare molecular ions in chosen states via probabilistic measurements.\cite{chouNature2017,najafianNatureComm2020,liuScience2024} These quantum-logic protocols are currently suitable for a single molecule at a time.

Optical pumping can be applied to larger ensembles and has been used to enhance the rotational ground state population in translationally and vibrationally cold HD$^+$ and MgH$^+$ ions.\cite{schneider2010all,staanum2010rotational} Each of these relied on driving vibrational transitions within the ground electronic state, and such techniques are limited to polar molecules. Optical pumping via excited electronic states can be feasible for molecules with diagonal Franck--Condon factors, where vibrational branching is less of a concern.\cite{lienNatureComm2014}

Optical preparation of a chosen state in a neutral molecule can restrict the states of a subsequently produced ion. The restriction may be based on available energy, Franck--Condon overlap of vibrational state,\cite{prattRPP1995} or parity and rotational propensity or selection rules.\cite{xieJCP1990} For example, optical preparation of autoionizing states of HfF is the starting point for trapping HfF$^+$ for use in a search for the electron electric dipole moment.\cite{lohJCP2011} Resonance-enhanced multi-photon ionization (REMPI) has been used to load ion traps with state-selected N$_2^+$,\cite{tong2010sympathetic} H$_2^+$,\cite{schmidtPRAppl2020} SiO$^+$,\cite{stollenwerkJMS2019} and HD$^+$,\cite{zhangPRA2023} among other species.

In this paper, we demonstrate the use of $(2+1)$ REMPI to load O\textsubscript{2}\textsuperscript{+} in an ion trap, where it is translationally cooled to a Coulomb crystal by co-trapped atomic ions. The oxygen molecular cation plays a significant role in gas-phase reactions in the upper atmosphere~\cite{schunkRevGeophysSpacePhys1980,deng2012rotational} and has potential applications as an optical clock and for tests of fundamental physics.\cite{carolloAtoms2018,hannekeQST2021,wolfNJP2024} We also ionize O$_2$ in a molecular beam to measure the spectrum of two-photon transitions between the ground vibrational states of the neutral molecule's $X\,^3\Sigma_g^-$ and $d\,^1\Pi_g$ states. We report spectroscopic parameters of the $d$ state that are in agreement with those in ref.~\onlinecite{lewis1999perturbations} and significantly shifted from those in refs.~\onlinecite{surJCP1991,mccann1993two}.

We photoionize O$_2$ through the  rotationally resolved Rydberg state $d\,^1\Pi_g$, which has molecular orbital configuration $3s\sigma_g$.  The potentials of Rydberg states often show near perfect Franck--Condon overlap with those of the ionic ground state,\cite{prattRPP1995} thus giving vibrational selectivity. One-color $(2+1)$ REMPI through the $d$ state's $v^\prime=0$ level requires radiation of wavelength approximately 301~nm. Here, $v$ is the vibrational quantum number, and we use $^{\prime\prime}$, $^\prime$, and $^+$ to denote the neutral ground, neutral excited, and ionic ground electronic states. The ion's rotational distribution is determined by selection and propensity rules among the angular momentum of the ion, photoelectron, and photon. We discuss strategies for maximizing the population in the ion's ground rovibrational state.

\begin{figure*}
\includegraphics[width=0.75\linewidth]{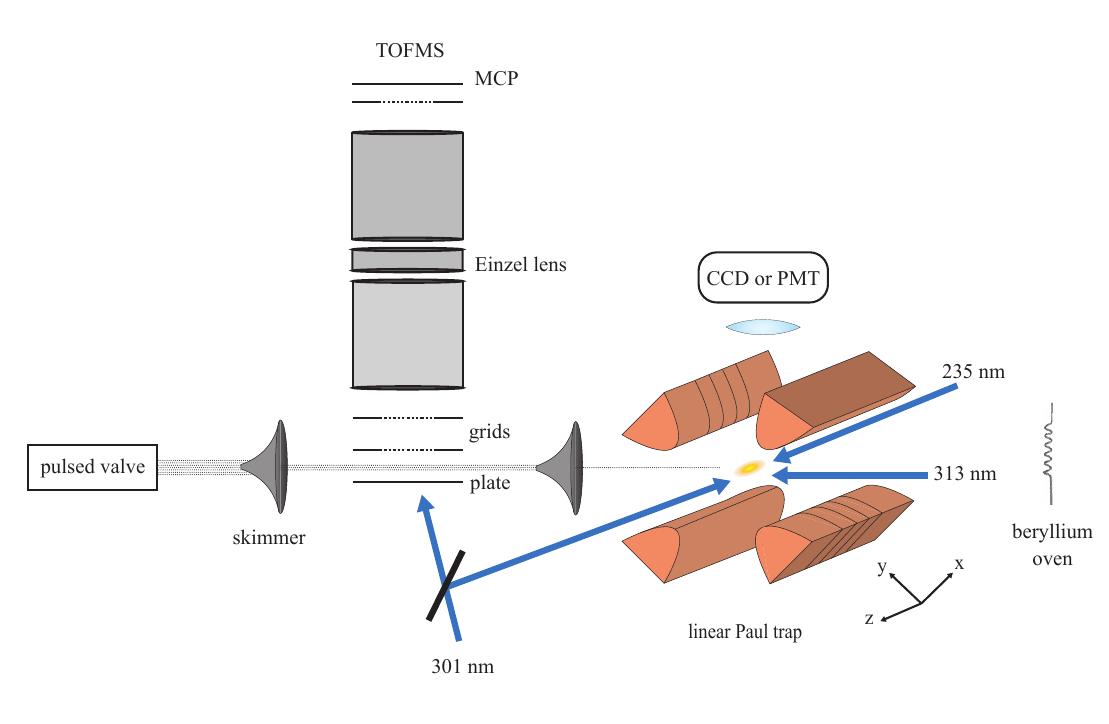}
\caption{\label{fig:apparatus}Schematic of the apparatus. A pulsed, skimmed beam of oxygen passes through a time-of-flight mass spectrometer (TOFMS) and an ion trap. The oxygen REMPI laser (301~nm) may be directed to either location. The ion trap is equipped for co-trapping beryllium atomic ions, with associated lasers for loading (235~nm) and cooling/detection (313~nm).}  
\end{figure*}

\section{Experimental Setup\protect\\ }
Our apparatus includes a linear radiofrequency (Paul) ion trap and a time-of-flight mass spectrometer (TOFMS), both housed in an ultrahigh-vacuum chamber. Fig.~\ref{fig:apparatus} gives a schematic overview. A pulsed molecular beam traverses both the TOFMS and ion trap, and the REMPI laser can be directed to either location. Separate experiments were conducted to record the O$_2$ REMPI spectrum using the TOFMS and to load and translationally cool O$_2^+$ in the Paul trap.

\subsection{REMPI in TOFMS}
The $(2+1)$ REMPI spectrum of the O\textsubscript{2} molecule was obtained by ionizing a pulsed supersonic molecular beam\cite{scolesBook1988} within the TOFMS. The neutral oxygen is introduced to the chamber through a pulsed solenoid valve (Parker Series 9, PTFE poppet). We have used both 100\,\% oxygen and a mix of 5\,\% oxygen and 95\,\% argon, with absolute gas pressures approximately $2.5\times10^5~{\rm Pa}$. A beam skimmer (Beam Dynamics, Inc., 1.0-mm aperture) is located 66~mm beyond the valve and both selects the central part of the molecular beam and allows for differential pumping of the TOFMS relative to the beam source chamber. The center of the TOFMS is 88~mm beyond the skimmer aperture. Turbomolecular pumps maintain the vacuum on both sides of the skimmer. The solenoid valve is electrically activated for 125~$\mu$s, beginning 600~$\mu$s prior to the laser pulse. This results in a gas pulse of duration approximately 225~$\mu$s, full-width at half-maximum.

Tunable UV radiation at wavelengths around 301~nm ionizes O$_2$ through the $d\,^1\Pi_g$ Rydberg state. An Nd:YAG-pumped, pulsed dye laser generates radiation at twice this wavelength, which is frequency-doubled in a BBO crystal. The pulsed laser uses a mix of Rhodamine 610 and 640 dyes in methanol. The laser fires at a rate of 10~Hz, with pulse duration approximately 7~ns and bandwidth of approximately $0.09~{\rm cm}^{-1}= 2.8~\textrm{GHz}$ at the fundamental. A small amount of the fundamental pulse is picked off and directed to commercial wavelength meter (HighFinesse WS7-60), which has an absolute accuracy of 150~MHz. As the laser steps across the REMPI spectrum, we measure the laser frequency at every step. Throughout the scan, we adjust the BBO crystal angle to maintain phase-matching and thus UV energy. We pick off a small amount of the UV light, direct it to a photodiode, and record the average energy at every frequency step. The linearly polarized pulse, of energy approximately 200~$\mu$J, is focused with a lens of focal length 150~mm. 

The TOFMS consists of one plate and two grid electrodes in a Wiley--McLaren configuration,\cite{wileyRSI1955} a flight region that is free of electric fields except for an embedded Einzel lens,\cite{adamsJPE1972} and a microchannel plate detector.\cite{kuzhanThesis2021} The plate and grid electrodes are held at ground during the laser pulse. After a delay of 1~$\mu$s, they are energized within $\lesssim100$~ns. The microchannel plate detector's signal is amplified and digitized at 2~GS/s. We integrate only the signal around the expected arrival time of O$_2^+$ ions.

\subsection{REMPI in Trap}\label{sec:expTrap}

The same molecular beam passes through a second skimmer (Beam Dynamics, Inc., 0.5~mm aperture), located 123~mm beyond the first skimmer. The ion trap is 267~mm beyond the second skimmer. The trap vacuum chamber is pumped with a combination non-evaporable getter and ion pump. It takes the gas pulse an additional 600~$\mu$s to arrive at the trap. When ionizing in the trap, we typically use laser energy of about 1 mJ per pulse, focused with a lens of 150~mm focal length.

    The linear ion trap consists of OFHC copper electrodes mounted with titanium dowels and screws into MACOR spacers. The ends of the electrodes closest to the ions are curved to approximate a two-dimensional quadrupolar equipotential. The radial distance from the ions to the electrodes is $r_0 = 1.25~\textrm{mm}$. One pair of opposing electrodes are driven with potential $V_0\cos(\Omega t)$, while the other pair are driven $180^\circ$ out of phase. A counter-wound transformer creates the two phases of the rf.\cite{henshonThesis2023} The trap electrodes and transformer together form an $LC$ circuit that is driven on resonance. Typical parameters are $V_0 = 30$~V and $\Omega = 2\pi(11~\textrm{MHz})$. For $^{16}\textrm{O}_2^+$, this produces a Mathieu parameter\cite{paulRMP1990}
    \begin{equation}
    	q = \frac{2e(2V_0)}{m r_0^2\Omega^2}     
    \end{equation}
     of $0.05$ and a radial secular frequency in the pseudopotential approximation\cite{raizenPRA1992} 
    \begin{equation}
    	\omega_r = \frac{q\Omega}{2\sqrt{2}}
    \end{equation}
    of $2\pi(190~\textrm{kHz})$. Here, $e$ is the magnitude of the ion's charge and $m$ is its mass. Note that the radial frequency scales inversely with mass; this scaling plays an important role in the identification of our trapped ion species.

To ensure axial confinement, two opposing electrodes are divided into five segments each with independently controllable potentials that can be added as a DC offset to the trap rf. The central three segments are each of length $2z_0 = 3.0~\textrm{mm}$, and our trap has a geometric factor\cite{raizenPRA1992} $\kappa \approx 0.14$. Typical parameters have the central segment held at 0~V with adjacent segments at $U = 2.0$~V. With these parameters, the axial frequency
\begin{equation}
	\omega_z = \sqrt{\frac{2\kappa e U}{m z_0^2}}
\end{equation}
for O$_2^+$ is $2\pi(140~\textrm{kHz})$. The axial confinement modifies the radial potential such that the true radial frequency is
\begin{equation}
 \omega_{x,y} = \sqrt{\omega_r^2-\omega_z^2/2}, \label{eq:xyfreq}
\end{equation}
which for O$_2^+$ is $2\pi(160~\textrm{kHz})$. With the central segment at 0~V the $x$ and $y$ frequencies are approximately the same. This degeneracy may be broken by moving the central segment voltage away from ground.

We can apply an independent static voltage to each of the 12 electrodes (10 segments and 2 unsegmented), such that we can null many stray electric fields and position the ions on the radiofrequency null. Several electrodes are also wired for the application of an oscillatory potential in the frequency range of the ions' motional frequencies. One long electrode allows for pushing along $x$, one of the central segements allows for pushing along $y$, and one of the off-center segments creates an electric field with a component along $z$. We can use the potentials to resonantly excite the ions along each trap axis.

    \subsection{\label{sec:level2}Sympathetic cooling and resonant mass detection}
    We sympathetically cool O$_2^+$ with co-trapped beryllium atomic ions, $^9\textrm{Be}^+$, which have a Mathieu $q$ parameter of 0.17 for the typical parameters above.  Our choice of $^9\textrm{Be}^+$ is motivated primarily because our lab already possesses the relevant cooling and photoionization lasers. Other coolant ions such as $\textrm{Mg}^+$ and $\textrm{Ca}^+$ are possible. Because of the mass-dependent radial pseudopotential, ions in a Coulomb crystal are typically arranged with the heavier ions surrounding the lighter ones. For $\textrm{O}_2^+$, that means $\textrm{Be}^+$ and $\textrm{Mg}^+$ would form the core of a crystal surrounded by $\textrm{O}_2^+$, wereas the locations would be inverted for $\textrm{Ca}^+$ and other heavier coolants. A closer mass-match tends to increase the collisional coupling between the ion species, whereas differences in mass separate their motional resonances in fluorescence mass spectrometry.
 
In our apparatus, an electric current heats a beryllium-wrapped tungsten wire to increase the presence of neutral beryllium in the trapping region. We ionize this beryllium via $(1+1)$~REMPI with linearly polarized radiation near 235~nm. The resonant step is $2s\,^1S_0$ to $2p\,^1P_1$. The laser light is produced by frequency-doubling 470~nm light in a BBO crystal. The 470~nm light is produced by a commercial external cavity diode laser (ECDL). The BBO crystal is held in a monolithic bowtie power-enhancement cavity.~\cite{plucharThesis2018} Typical UV powers are several hundred microwatts.
 
 \begin{figure}
\includegraphics[width=\linewidth]{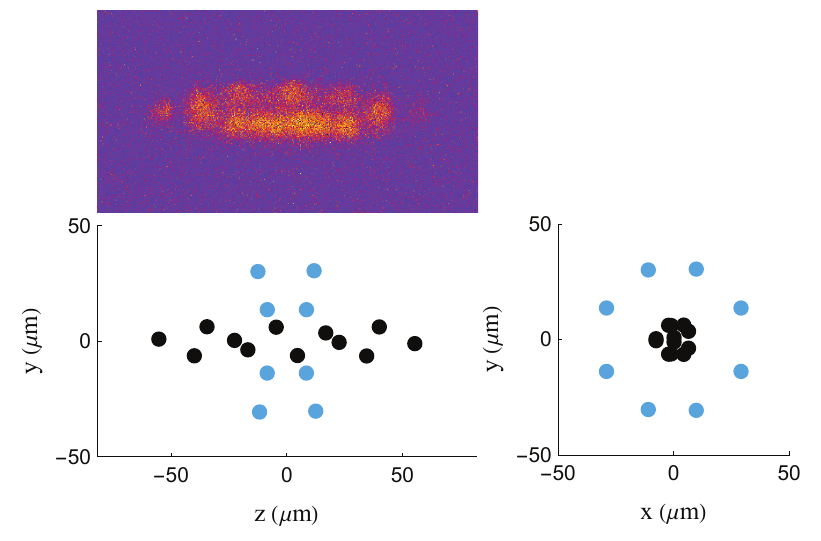}
\caption{\label{fig:crystal} EMCCD image (top) and simulation (bottom) of an ion crystal with Be$^+$ and O$_2^+$. The image has the same scale as the simulation. Only the beryllium ions fluoresce. The simulation has 12 Be$^+$ ions (black) and 8 O$_2^+$ ions (blue) and uses the trap parameters from the main text. There is not enough information in the EMCCD image to provide number resolution, so the simulation is included for its qualitative similarity to the image. The presence of oxygen was verified for this crystal by use of resonant mass detection.}
\end{figure}
 
    We laser-cool the Be$^+$ on a cycling transition between $2s\,^2S_{1/2}, F=2,$ and $2p\,^2P_{3/2}, F=3,$ at 313~nm. UV radiation with powers of order milliwatts is generated by third-harmonic generation of an amplified ECDL at 939~nm.\cite{carolloOE2017} The cooling laser is circularly polarized. Its $k$-vector has projections on all three trap axes, ensuring comprehensive cooling of the ions. Prior to admitting the oxygen gas pulse, we typically load a Coulomb crystal consisting of 10--20 Be$^+$ ions. Figure~\ref{fig:crystal} shows an example of a multi-species ion crystal.
   
The cooling laser's frequency is stabilized against drift by referencing it to a stabilized helium--neon laser via a scanning Fabry--Perot cavity.~\cite{carolloOE2017} Acousto-optic modulators enable frequency shifts of the UV beam. They also allow the laser to be turned on and off with sub-microsecond timing. We apply lasers with frequencies either near resonance or detuned approximately 400~MHz below resonance. 
   
Scattered light from the cooling laser serves as our primary detection channel for experiments in the ion trap. A microscope ($\textrm{NA}\approx 0.4$, magnification 30) can direct the light to either an EMCCD camera or a photomultiplier tube. The Be$^+$ scattering rate depends on the laser detuning, any Doppler shift from the ion's motion, and any larger motion that takes the ions out of the beam's focus.

We detect the charge-to-mass ratio of ions in the trap with fluorescence mass-spectrometry.\cite{babaJJAP1996,rothPRA2007}  An oscillatory potential on one of the electrodes resonantly excites the radial motion of one ion species (eq.~\ref{eq:xyfreq}). Since a typical ion in our trap is singly charged, the frequency is a proxy for mass. The excited ions collisionally excite all other ions in the trap, including the Be$^+$ ions. This leads to a change in the beryllium fluorescence that depends on the frequency of the applied electric field. Depending on the laser detuning, the degree of excitation, and other parameters, the fluorescence can either increase or decrease when the ions are excited.

After loading ions, we repeat the following experimental sequence to probe the radial frequencies of the trapped ions, and thus their mass. First, we cool the {Be$^+$} ions with the detuned laser for 5~ms. The beryllium sympathetically cools any other ions, forming a Coulomb crystal. Second, with all lasers off, we apply the oscillatory potential to one of the central segmented electrodes with an amplitude of order 100~$\mu$V for duration 2~ms. Third, we apply the laser tuned near resonance for duration 1~ms and count photons received by the photomultiplier tube. We repeat this sequence 25 times for each frequency. By returning to the detuned cooling each time, each experiment is independent of the ones preceding it.

\section{Results and Discussion\protect\\ }

\subsection{Spectroscopy}

\begin{figure}
\includegraphics[width=\linewidth]{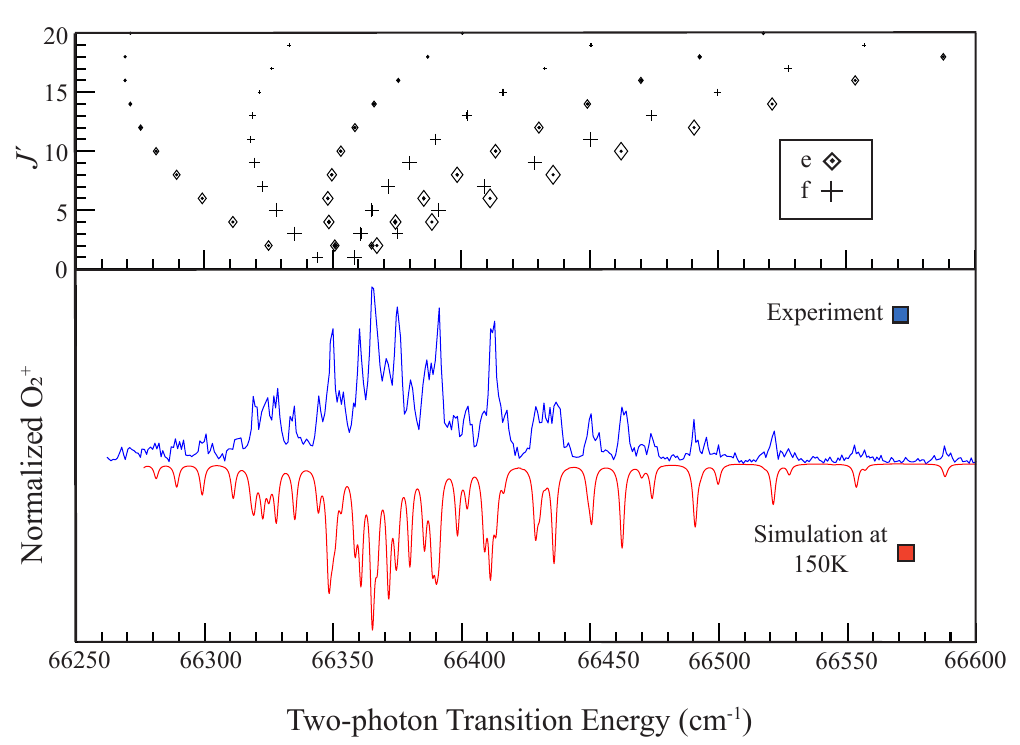}
\caption{\label{fig:broadSpectrum} $(2+1)$~REMPI spectrum of a pure oxygen beam. Experimental data (blue, upright) match well with simulation (red) with temperature 150~K. The upper panel shows seven rotational branches; left to right, they are N, O, P, Q, R, S, T corresponding to $\Delta N = -3$ through $+3$. The symbol size indicates the relative transition strength at 150~K. The labels $J^\prime$ and $e/f$ correspond to the $d\,^1\Pi_g$ state.}
\end{figure}

\begin{figure}
\includegraphics[width=\linewidth]{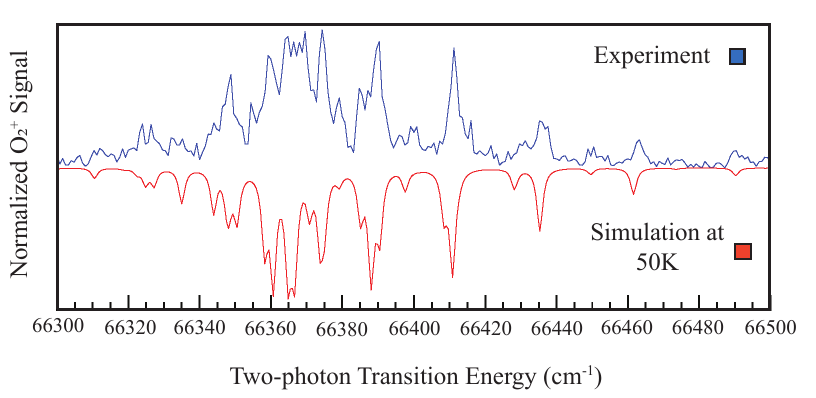}
\caption{\label{fig:50K}Experimental $(2+1)$ REMPI spectrum (blue) of a beam of 5\,\% O$_2$ and 95\,\% Ar, overlaid with a simulation (red) with a temperature of 50~K.}
\end{figure}

Figure~\ref{fig:broadSpectrum} shows the $(2+1)$~REMPI spectrum of O\textsubscript{2} in the $v^{\prime\prime}=v^\prime=0$ band of the  $X\,^3\Sigma_g^-\rightarrow\rightarrow d\,^1\Pi_g$ transition. The bottom panel shows experimental data upright in blue from a gas beam of pure oxygen. The data shown is energy-normalized by taking the molecular ion signal at a given wavenumber and dividing it by the UV photodiode signal. 
Inverted in red is a simulated spectrum that is generated by the program PGOPHER\cite{westernJQSRT2016} with a temperature of 150~K. This temperature was determined by comparing the relative intensities of the transitions in the data and in simulation; we estimate an uncertainty of $\pm 40$~K. The simulation includes a fixed Lorentzian linewidth (full-width at half-maximum) of $2~\textrm{cm}^{-1}$, which accurately matches our experimental linewidth and is consistent with prior results.\cite{surJCP1991,lewis1999perturbations} At higher laser energies, we see significant power broadening and saturation. Fig.~\ref{fig:50K} displays data from a beam of 5\,\% oxygen and 95\,\% argon, along with a simulated spectrum with temperature 50~K. We estimate an uncertainty of $\pm 20$~K on this temperature.

The top panel of Fig.~\ref{fig:broadSpectrum} shows Fortrat parabolas, which appear in seven branches. Two-photon transitions allow $J^\prime - J^{\prime\prime} = 0, \pm1, \pm2$. Because the ground-state triplet splittings are finer than the transition linewidth and the excited state is a singlet, the transition energy is best characterized by $N^\prime - N^{\prime\prime} = -3$ through $3$, corresponding to rotational branches N, O, P, Q, R, S, T. Since we are most interested in the rotational state from which we ionize, the figure plots the transitions with respect to $J^\prime=N^\prime$. They are labeled by their $\Lambda$ doublet component $e/f$. Oxygen nuclei are symmetric, so the $d$ state's gerade symmetry means its rotation states are even parity,\cite{herzbergVolI1950} such that odd $J^\prime$ have only the $f$ state and even $J^\prime$ have only the $e$ state.\cite{brown_carrington}

We model the spectra with the Hamiltonian
\begin{equation}
	\hat{H}_{\rm eff} = \hat{H}_{\rm el,vib}+\hat{H}_{\rm rot}+ \hat{H}_{\rm SS} + \hat{H}_{\rm SR}.
\end{equation}
Here, $\hat{H}_{\rm el,vib}$ is the electronic and vibrational Hamiltonian, $\hat{H}_{\rm rot}$ accounts for rotation, and $\hat{H}_{\rm SS}$ and $\hat{H}_{\rm SR}$ account for spin--spin and spin--rotation interactions, where applicable. Rotational ($B_0$, $D_0$), spin--spin ($\lambda$, $\lambda_D$), and spin--rotation ($\gamma$) parameters of the $X\,^3\Sigma_g^-$ state are well known from extensive previous spectroscopy experiments. Here, we use the parameters in Drouin \emph{et al}.,\cite{drouinJQSRT2010} except we correct the sign of the spin--spin coupling parameter $\lambda$ and its distortion correction $\lambda_d$ to be in agreement with Tretyakov \emph{et al.}\cite{tretyakov200560} The $d\,^1\Pi_g$ Hamiltonian has no spin terms. We fit the spectra to determine the $d$ state's band origin $T_{00}$ and the rotational constant $B_0$. We fix the centrifugal distortion term $D_0$ to zero; its inclusion does not improve the quality of the fit with our relatively low values of $J^\prime$. We use the program PGOPHER for the fit; ref.~\onlinecite{westernJQSRT2016} has a full discussion of the Hamiltonian parameterization and fit routine. Our fit includes 27 rotational transitions with quantum numbers as high as $J^\prime=N^\prime=18$. We observe a standard deviation of the fit residuals of $0.5~\textrm{cm}^{-1}$.

Table~\ref{tab:constants} compares our results for the $d\,^1\Pi_g$ state with those from other experiments. Table~\ref{tab:groundConstants} lists constants used for the $X\,^3\Sigma_g^-$ state. Table~\ref{tab:transitions} gives the list of the transitions included in the fit.  Our rotational constant $B_0 = 1.6798(12)~\textrm{cm}^{-1}$ is in agreement with that determined by refs.~\onlinecite{sur1985optical,surJCP1991,lewis1999perturbations}. It matches the rotational constant of the ion's ground electronic state (1.67996(26)~cm$^{-1}$),\cite{coxonJMS1984} as expected for a neutral Rydberg state. Our value for the band origin $T_{00} = 66357.33(15)~\textrm{cm}^{-1}$ agrees with the result of Lewis~\emph{et al}.,\cite{lewis1999perturbations} which analyzed data presented by Ogorzalek-Loo.\cite{ogorzalek1989multiphoton,ogorzalek1989multiphotonpaper} Earlier work by Sur~\emph{et al.}\cite{surJCP1991} and McCann~\emph{et al.}\cite{mccann1993two} disagree by tens of cm$^{-1}$. We suspect that our laser's absolute calibration has a smaller uncertainty, though neither reference discusses their wavelength measurement procedure.
Our fit of the 50-K spectrum yields values consistent with our results, but with larger uncertainties ($1.1~\textrm{cm}^{-1}$) because we observe fewer lines at the lower temperature.

\begin{table}
\caption{\label{tab:constants}Comparison of our experimentally determined spectroscopic parameters with other relevant literature for the  $v^\prime=0$, $d\,^1\Pi_g$ state of O$_2$.}
\begin{ruledtabular}
\begin{tabular}{ccccc}
 Constants    &    This work    
     &  $\vcenter{\hbox{\strut Lewis \emph{et al.}}\hbox{\strut~~(1999)\cite{lewis1999perturbations}}}$
     &  $\vcenter{\hbox{\strut McCann \emph{et al.}}\hbox{\strut~~(1993)\cite{mccann1993two}}}$
     &  $\vcenter{\hbox{\strut Sur \emph{et al.}}\hbox{\strut~~(1991)\cite{surJCP1991}}}$  \\ \hline \\
 $T_{00}$  &  $66357.33(15)$  &  $66358$  &  $66409$\footnote{This is reported as the transition energy for Q(1).}&$66380.15$\\
 $B_0$  &  $1.6798(12)$  &  $1.682$  &  $-$  &  $1.68$\\
 $D_0$  &  $0$\footnote{fixed}  &  $7.7\times10^{-6}$  &  $-$  &  $4.34\times10^{-6}$
\end{tabular}
\end{ruledtabular}
\end{table}

\subsection{State Selectivity}

Because we can resolve the rovibrational structure of the two-photon transition in the neutral, we can spectroscopically prepare the ion's quantum state. The ion's ground electronic state $X\,^2\Pi_g$ is of type Hund's case (a) with spin-orbit splitting $A = 200.289(20)~\textrm{cm}^{-1}$ and rotational constant $B = 1.67996(26)~\textrm{cm}^{-1}$ in its ground vibrational state.\cite{coxonJMS1984} Our goal is to produce a high percentage in the ground rovibrational state $|v^+=0, \Omega^+=\tfrac{1}{2}, J^+=\tfrac{1}{2}\rangle$. The $(2+1)$ REMPI photons do not have sufficient energy to populate an excited electronic state of the ion, so all ions are in the ground $X$ state. Although the ionizing photon has enough energy to populate the $v^+=1$ state, the Franck--Condon overlap of the Rydberg $d$ state and ion ground $X$ state is near perfect. Perturbing valence states in the neutral could disrupt that overlap, but photoelectron spectroscopy shows a 91\,\% probability of populating $v^+=0$.\cite{stephensJCP1990}

Even if the two-photon transition populates a single rotational level in the $d$ state, the ion can still be produced in one of several rotational states. The gerade symmetry of our Rydberg and ionic states restricts the quantum number of the orbital angular momentum transfer to the ion to be an even integer.\cite{xieJCP1990} If the orbital angular momentum transfer is zero, then the total angular momentum change at the ionization step (including the photoelectron's spin) is one of $J^+-J^\prime = \pm\tfrac{1}{2}$. Following the calculations in refs.~\onlinecite{braunsteinJCP1992,germannJCP2016}, we find that ionizing from the lowest rotational state $J^\prime = 1$ will yield a rotational distribution in the molecular ion of $\tfrac{1}{3}$ in $|\Omega^+=\tfrac{1}{2}, J^+=\tfrac{1}{2}\rangle$, $\tfrac{1}{6}$ in $|\Omega^+=\tfrac{1}{2}, J^+=\tfrac{3}{2}\rangle$, and $\tfrac{1}{2}$ in $|\Omega^+=\tfrac{3}{2}, J^+=\tfrac{3}{2}\rangle$.  
The uncertainty in these rotational branching fractions comes from the possibility of orbital angular momentum transfer of 2 or more. Such higher-order transfers typically contribute less to the overall ionization probability, but would allow higher $J^+$ to be populated. For example, an orbital angular momentum transfer of 2 would allow $|J^+-J^\prime| \le \tfrac{5}{2}$. If this higher orbital angular momentum transfer of 2 had 25\,\% of the probability of the zero-orbital-angular-momentum transfer, then any molecules ionized from $J^\prime = 1$ would be end up in the $|\Omega^+=\tfrac{1}{2}, J^+=\tfrac{1}{2}\rangle$ state 29\,\% of the time instead of 33\,\%.

A second laser at 323~nm could enhance the probability of populating $|\Omega^+=\tfrac{1}{2}, J^+=\tfrac{1}{2}\rangle$ if the laser is tuned above the ionization threshold but below the energy of $|\Omega^+=\tfrac{1}{2}, J^+=\tfrac{3}{2}\rangle$. The intensity of this second laser should be much higher than that of the 301~nm laser, such that the $(2+1^\prime)$ REMPI rate exceeds the $(2+1)$ rate. The trap's electric field can shift the ionization potential in ways that can impact the state-preparation fidelity in such threshold schemes.\cite{blackburnSciRep2020,shlykovAdvQuantTech2023,zhangPRA2023} For larger, weaker traps with smaller electric fields, the ground-state fidelity can still be in excess of 90\,\%.\cite{tong2010sympathetic}

\begin{figure}
\includegraphics[width=\linewidth]{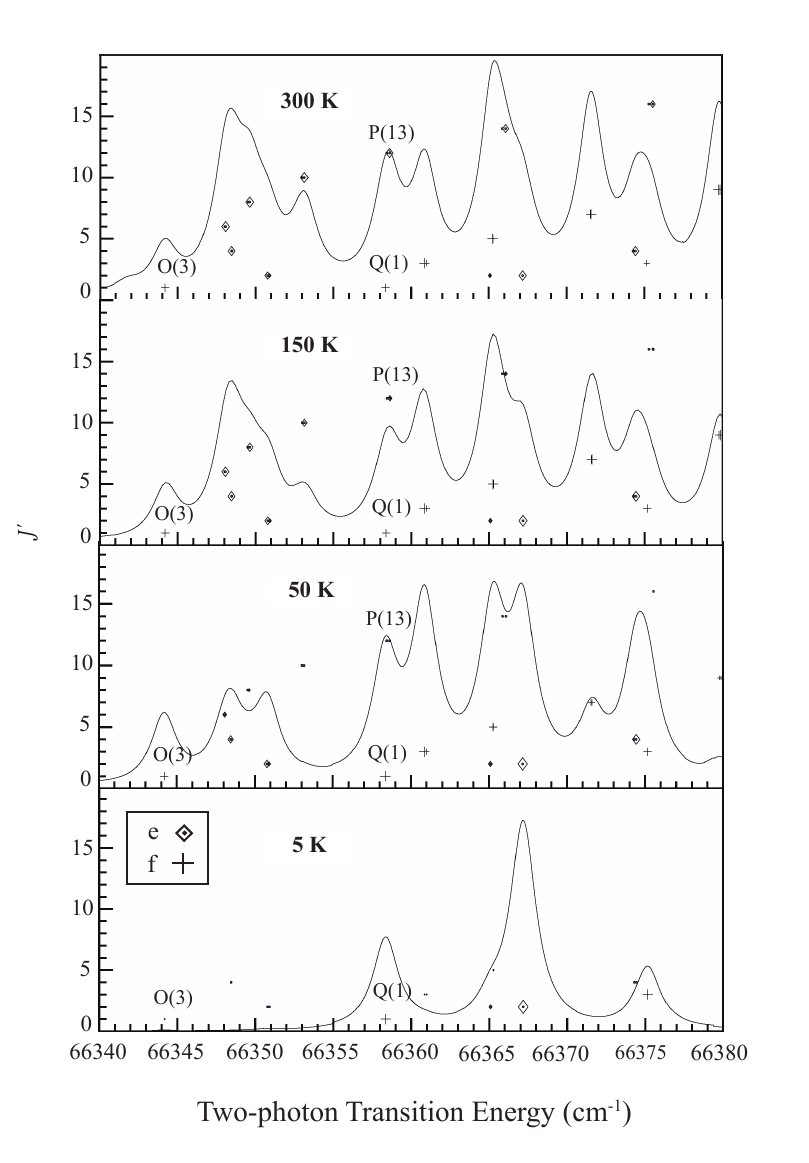}
\caption{\label{fig:narrowSpectrumAll} Calculated $(2+1)$ REMPI spectrum with overlaid Fortrat parabolas, for temperatures of 300~K, 150~K, 50~K, and 5~K. The range shown includes the Q(1) and O(3) transitions, which are the best suited for producing ions in the lowest rotational state. Transitions with normalized intensities less than $10^{-6}$ are omitted.}
\end{figure}

There are exactly two two-photon transitions that excite the $|d\,^1\Pi_g, N^\prime = 1, J^\prime = 1\rangle$ state. We observe the O(3) transition, from $|X\,^3\Sigma_g^-, N^{\prime\prime} = 3, J^{\prime\prime} = 2\rangle$, to be at 66344.56~cm$^{-1}$, and the Q(1) transition, from $|X\,^3\Sigma_g^-, N^{\prime\prime} = 1, J^{\prime\prime} = 2\rangle$, to be at 66358.18~cm$^{-1}$. The choice of which transition to use depends on the temperature of the oxygen. At low temperatures, the Q(1) transition has much higher intensity because the $N^{\prime\prime}=1$ state has higher occupation than $N^{\prime\prime}=3$. At higher temperatures, the Q(1) transition overlaps with the P(13) transition to $|d\,^1\Pi_g, N^\prime = 12, J^\prime = 12\rangle$, which we calculate to be only 0.05~cm$^{-1}$ below Q(1) and thus not possible to resolve. At these higher temperatures, the O(3) transition remains free of any overlapping transitions and is the preferred choice. Figure~\ref{fig:narrowSpectrumAll} plots the calculated REMPI spectrum at several temperatures, along with Fortrat parabolas indicating transition locations. At 150~K, the Q(1) transition is 33\,\% more intense than O(3) and still three times stronger than P(13). By 50~K, Q(1) is several hundred times stronger than P(13) and twice as strong as O(3), making Q(1) the better choice. At 5~K, as others have seen with pulsed supersonic beams,\cite{surJCP1991} both Q(1) and O(3) are unobstructed, but Q(1) is much higher intensity, making it the better choice. At 300~K, P(13), Q(1), and O(3) are all comparable in intensity; since O(3) is unobstructed, it is the better choice.

An alternative Rydberg state is $C\,^3\Pi_g\,(v^\prime=0)$, which also has molecular orbital configuration $3s\sigma_g$ and for which $(2+1)$~REMPI  occurs in the range 302-305 nm.\cite{surJCP1986} This state would ionize exclusively to the $v^+=0$ state because the ionizing photon does not have enough energy to populate vibrationally excited states. Nonetheless, the two-photon transition is not rotationally resolved,~\cite{surJCP1986} which impedes rotational selectivity. Although transitions through the $C$ state's $v^\prime = 2$ level are rotationally resolved~\cite{surJCP1986} -- and have been used to load O$_2^+$ in an ion trap\cite{zagorecmarksFaradayDiscuss2024} -- the state shows significant non-Franck--Condon behavior in ionization and should not be expected to produce ions in a single vibrational state. Indeed, photoelectron spectroscopy of both the $d$ and $C$ states shows significant non-Franck--Condon behavior in ionization for $v^\prime \ge 1$.\cite{katsumata1986,stephensPhysScr1990} Many higher Rydberg states of O$_2$ have been identified and studied.\cite{katsumataAppSpectRev1992} Some have photoelectron spectra suggesting a higher fraction of transitions with $v^+=v^\prime$ for $v^\prime \ge 1$.\cite{parkJCP1988,parkJCP1988a} A higher Rydberg state has been used to load O$_2^+$ in an ion trap from room-temperature background gas; in that case, collisions with the neutral gas thermalized the ions to the ground vibrational state and a room-temperature distribution of rotational states.\cite{dorflerNatureComm2019}

\subsection{Trapping}

\begin{figure}
\includegraphics[width=\linewidth]{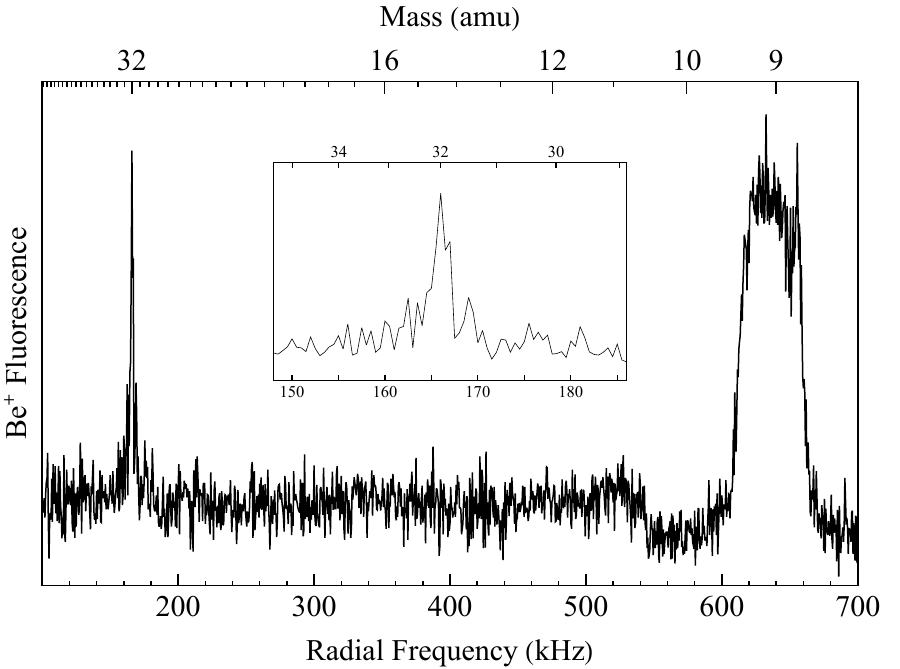}
\caption{\label{fig:massSpectrum} Fluorescence mass spectrometry scan showing the presence of both Be$^+$ at 640~kHz and O$_2^+$ at 166~kHz. The inset expands the frequency axis near the O$_2^+$ resonance.}
\end{figure}

Fig.~\ref{fig:massSpectrum} demonstrates the trapping of photoionized O$_2^+$ ions. It shows the fluorescence of beryllium ions versus the frequency of the radial excitation drive, with strong responses near the expected frequencies for ions of mass 32~amu (near 160~kHz) and 9~amu (near 640~kHz), corresponding to $^{16}\textrm{O}_2^+$ and $^9\textrm{Be}^+$. Both ions were loaded via photoionization. The neutral oxygen was ionized from a beam of pure oxygen using the Q(1) resonance. The stronger beryllium response may be because the beryllium excitation affects the fluorescence directly, while the oxygen affects the fluorescence only through collisions. We note that the resonances are of comparable width in terms of mass.

As a check on the procedure, we loaded the two species sequentially. We first photoionized the beryllium ions and verified that there was no response to modulation in the range 100--220~kHz. The response at 166~kHz only appears after applying both the oxygen REMPI laser and the gas pulse at the same time.

As a check that the oxygen REMPI remains rotationally resolved and unsaturated, we attempted loading at several different laser frequencies. No oxygen ions were loaded at the two-photon energy 66340.2~cm$^{-1}$, which is a low point in the spectrum just below the O(3) resonance.

Fluorescence mass spectrometry is nondestructive in that it only disturbs the motion of the molecular ions; they remain trapped, and -- except for the fluorescing ion -- their internal states are not directly disturbed. When combined with state-selective loading, there are broad applications in experiments requiring quantum state control and benefiting from the long storage times of Paul traps. For example, it can be used to investigate products of state controlled chemical reactions, photodissociation, and photoionization without needing to reload the trap between experiments.\cite{drewsen2004nondestructive} In cases where reaction rates are slow compared to the millisecond-scale detection time, the technique allows for real-time monitoring of reaction products. Ion traps are increasingly becoming a common technique to probe chemical reactions,\cite{ratschbacher2012controlling,dorflerNatureComm2019} analyze their products,\cite{zagorecmarksFaradayDiscuss2024,campbell2020spectroscopy} and study ions that are common in interstellar environments.\cite{brunken2014h2d+,catani2020translationally} The nondestructive detection approach can aid these studies in identification of ionic species. 

This demonstration with O$_2^+$ is an enabling step towards the development of optical clocks based on molecular vibration. In addition to their usefulness as time and frequency references, such clocks have the potential to probe new physics models of quantum gravity or dark matter.\cite{hannekeQST2021} Vibrational transitions in nonpolar molecules are electric-dipole forbidden, but may be driven as electric quadrupole or two-photon transitions. The narrowness of these transitions makes them amenable for precision spectroscopy, but also leads to low transition rates. For O$_2^+$, two-photon vibrational transitions may be driven from the ground vibrational state $v^+=0$ to excited states by use of lasers that have wavelengths ranging from 10.6~$\mu$m for $v^+=1$ to 418~nm for $v^+=38$ and beyond. To detect these transitions, we have proposed photodissociation of vibrationally excited molecules.\cite{carolloAtoms2018} Fluorescence mass spectrometry will allow us to probe for the resulting atomic ion O$^+$ without losing any remaining sample of O$_2^+$. The mass resolution shown in fig.~\ref{fig:massSpectrum} is sufficient to resolve a potential atomic signal at 16~amu.

\section{Conclusion\protect\\ }

In this paper, we have demonstrated the use of $(2+1)$ REMPI to prepare  O$_2^+$ in a linear Paul trap. We have used motional excitation as a nondestructive method to detect photoionized ions. A fit to our molecular-beam REMPI data clears up a discrepancy in reports of the enegy $T_{00}$ of the $d\,^1\Pi_g$ state of O$_2$. Simulations using our fit parameters show the best transitions for preparing state-selected O$_2^+$ at various temperatures. In particular, at low temperatures the Q(1) transition is the strongest for ionization from $J^\prime = 1$. At higher temperatures, Q(1) is obstructed by the P(13) transition, so O(3) becomes preferable. We suggest an extension to $(2+1^\prime)$ REMPI, where the ionization photon is at 323~nm, which should be capable of loading a higher fraction of the ions in the $|v^+=0,\Omega^+=\tfrac{1}{2}, J^+=\tfrac{1}{2}\rangle$ rovibrational state of the $X\,^2\Pi_g$ ground electronic state. Fluorescence mass spectrometry of O$_2^+$ with co-trapped Be$^+$ both demonstrates the successful photoionization of oxygen and enables further study of this molecule for chemistry or precision measurements.

\begin{acknowledgments}
This material is based upon work supported by the US National Science Foundation under grants RUI PM PHY-2207623 and RUI PHY-1806223.
\end{acknowledgments}

\section*{Author Declarations}
\subsection*{Conflict of Interest}
The authors have no conflicts to disclose.

\subsection*{Author Contributions}
\textbf{Ambesh Pratik Singh}: Formal Analysis (lead), Investigation (equal), Methodology (equal), Supervision (supporting), Visualization (equal), Writing/Original Draft Preparation (lead), Writing/Review \& Editing (equal). 
\textbf{Michael Mitchell}: Formal Analysis (supporting), Investigation (equal), Methodology (equal), Visualization (equal), Writing/Review \& Editing (equal). 
\textbf{Will Henshon}: Investigation (equal), Methodology (equal). 
\textbf{Addison Hartman} Investigation (equal), Methodology (equal). 
\textbf{Annika Lunstad}: Investigation (equal), Methodology (equal). 
\textbf{Boran Kuzhan}: Investigation (equal), Methodology (equal). 
\textbf{David Hanneke}: Conceptualization (lead), Formal Analysis (supporting), Funding Acquisition (lead), Investigation (equal), Methodology (equal), Supervision (lead), Visualization (equal), Writing/Original Draft Preparation (supporting), Writing/Review \& Editing (equal).

\section*{Data Availability}

The data that support the findings of this study are available from the corresponding author upon reasonable request.

\appendix*

\section{Additional tables}

\begin{table}[!h]
\caption{\label{tab:groundConstants}Spectroscopic parameters in cm$^{-1}$ of the  $v^{\prime\prime}=0$, $X\,^3\Sigma_g^{-}$ state of O$_2$, obtained from Drouin \emph{et al.} (2010),\cite{drouinJQSRT2010} and used in our fit.}
\begin{ruledtabular}
\begin{tabular}{ll}
 Constants    &    Drouin \emph{et al.} (2010)\cite{drouinJQSRT2010} \\ \hline 
 $B_{0}$&$1.4377$\\
 $D_0$&$4.841\times 10^{-6}$\\
 $\lambda$&$1.9848$\footnotemark[1]\\
 $\lambda_D$&$1.9\times10^{-6}$ \footnotemark[1]\\
 $\gamma$&$-0.0084$\\
 
\end{tabular}
\end{ruledtabular}
\footnotetext[1]{Negative sign removed for this constant from Drouin \emph{et al.} (2010),\cite{drouinJQSRT2010} in agreement with Tretyakov \emph{et al.} (2005).\cite{tretyakov200560} }
\end{table}

\begin{table}[!h]
\caption{\label{tab:transitions}Rotational transitions used for fitting the $0-0$ vibrational band of the two-photon $X\,^3\Sigma_g^- \rightarrow\rightarrow d\,^1\Pi_g$ transition of O$_2$.}

\begin{ruledtabular}
    \begin{tabular}{llllllll}
    \\
        $J^\prime$ & $N^\prime$ & $J^{\prime\prime}$ & $N^{\prime\prime}$ & \textbf{Freq. Obs.} & \textbf{obs-calc} & $\Delta N$ & $\Delta J$ \\ \hline
        9 & 9 & 10 & 11 & 66319.40 &-0.01& O & P \\ 
        5 & 5 & 6 & 7 & 66327.93 &0.06& O & P \\ 
        3 & 3 & 4 & 5 & 66334.38 &-0.66& O & P \\ 
        1 & 1 & 2 & 3 & 66344.56 &0.37& O & P \\ 
        8 & 8 & 10 & 9 & 66349.64 &0.01& P & O \\ 
        1 & 1 & 2 & 1 & 66358.18 &-0.18& Q & P \\ 
        3 & 3 & 4 & 3 & 66360.25 &-0.60 & Q & P \\ 
        3 & 3 & 2 & 3 & 66360.25 & -0.74 & Q & R \\ 
        5 & 5 & 6 & 5 & 66365.52 &0.27& Q & P \\ 
        5 & 5 & 4 & 5 & 66365.52 &0.25& Q & R \\ 
        7 & 7 & 8 & 7 & 66370.82 &-0.76& Q & P \\ 
        7 & 7 & 6 & 7 & 66370.82 &-0.72& Q & R \\ 
        4 & 4 & 2 & 3 & 66374.92 &0.50& R & S \\ 
        3 & 3 & 2 & 1 & 66374.92 &-0.23& S & R \\ 
        9 & 9 & 10 & 9 & 66380.25 &0.39& Q & P \\ 
        9 & 9 & 8 & 9 & 66380.25 &0.48& Q & R \\ 
        11 & 11 & 12 & 11 & 66391.01 &0.91& Q & P \\ 
        11 & 11 & 10 & 11 & 66391.01 &1.05& Q & R \\ 
        5 & 5 & 4 & 3 & 66391.01 &-0.08& S & R \\ 
        9 & 9 & 8 & 7 & 66428.95 &0.25& S & R \\ 
        8 & 8 & 6 & 5 & 66436.66 &0.86& T & S \\ 
        11 & 11 & 10 & 9 & 66450.29 &-0.12& S & R \\ 
        10 & 10 & 8 & 7 & 66462.30 & 0.01& T & S \\ 
        12 & 12 & 10 & 9 & 66490.28 &-0.45& T & S \\ 
        14 & 14 & 12 & 11 & 66521.23 &0.11& T & S \\ 
        16 & 16 & 14 & 13 & 66552.69 &-0.79& T & S \\ 
        18 & 18 & 16 & 15 & 66587.64 &-0.18& T & S \\ 
    \end{tabular}
\end{ruledtabular}
\end{table}

\providecommand{\noopsort}[1]{}\providecommand{\singleletter}[1]{#1}%

\end{document}